\documentclass{iau} 
\usepackage{graphicx}
\usepackage{natbib}

\title[Creating S0s with major mergers] 
{Creating lenticular galaxies with mergers}

\author[Miguel Querejeta et al.]   
{Miguel Querejeta$^{1}$,
  M.~Carmen Eliche-Moral$^{2}$,
  Trinidad Tapia$^{3}$,
  Alejandro Borlaff$^{4}$,
  Glenn van de Ven$^{1}$,
  Mariya Lyubenova$^{5}$, \\
  Marie Martig$^{1}$,
  Jes\'{u}s Falc\'{o}n-Barroso$^{4}$,
  Jairo M\'{e}ndez-Abreu$^{6}$, \\
  Jaime Zamorano$^{2}$,
 \and Jes\'{u}s Gallego$^{2}$}

\affiliation{$^{1}$ Max Planck Institute for Astronomy, K\"{o}nigstuhl 17, D-69117 Heidelberg, Germany\\
$^{2}$ Departamento de Astrof\'{i}sica, Universidad Complutense de Madrid, E-28040 Madrid, Spain\\
$^{3}$ Instituto de Astronom\'{i}a, UNAM, BC-22800 Ensenada, Mexico\\
$^{4}$ Instituto de Astrof\'{i}sica de Canarias, C/ V\'{i}a L\'{a}ctea, E-38200 La Laguna, Tenerife, Spain\\
$^{5}$ Kapteyn Institute, University of Groningen,  NL-9700 Groningen, The Netherlands\\
$^{6}$ School of Physics and Astronomy, University St Andrews, KY16 9SS, St Andrews, UK}

\pubyear{2016}
\volume{321}  
\jname{Outskirts of Galaxies}
\editors{A.~Gil de Paz, J.~Lee, \& J.~H.~Knapen, eds.}
\begin{document}

\maketitle

\begin{abstract}

Lenticular galaxies (S0s) represent the majority of early-type galaxies in the local Universe, but their formation channels are still poorly understood. While galaxy mergers are obvious pathways to suppress star formation and increase bulge sizes, the marked parallelism between spiral and lenticular galaxies (e.g.~photometric bulge--disc coupling) seemed to rule out a potential merger origin. Here, we summarise our recent work in which we have shown, through $N$-body numerical simulations, that disc-dominated lenticulars can emerge from major mergers of spiral galaxies, in good agreement with observational photometric scaling relations. Moreover, we show that mergers simultaneously increase the light concentration and reduce the angular momentum relative to their spiral progenitors. This explains the mismatch in angular momentum and concentration between spirals and lenticulars recently revealed by CALIFA observations, which is hard to reconcile with simple fading mechanisms (e.g.~ram-pressure stripping).
\end{abstract}

\firstsection 
\section{Introduction}

Contrary to their traditional position between ellipticals and spirals in morphological galaxy classification schemes, a number of independent studies have recently suggested that lenticular galaxies form a full sequence on their own, in parallel to spirals (S0a, S0b, etc.), which can be sorted based on the decreasing relevance of the spheroidal component in analogy to the sequence of spirals \citep{2010MNRAS.405.1089L,2011MNRAS.416.1680C,2012ApJS..198....2K}.
These parallel properties could well be revealing an  underlying evolutionary connection: are lenticulars indeed spiral galaxies which have undergone a certain transformation process? And, if so, did all lenticular galaxies originate through the same mechanism, or are they a ``mixed bag'' that results from very different evolutionary tracks, as suggested by \citet{1990ApJ...348...57V}?

One of the most popular processes to explain the origin of lenticular galaxies in the cluster regime is \textit{ram pressure stripping}. It consists on the expulsion of gas from a spiral galaxy falling into a cluster due to the pressure exerted by the intra-cluster medium. The lack of gaseous fuel results in the suppression of star formation, while discs can be preserved  \citep{2005AJ....130...65C,2006A&A...458..101A,2015MNRAS.447.1506M}. 

Nevertheless, observations imply that a large fraction of the local population of lenticular galaxies is not associated with the dense environment of clusters, but rather belongs to less densely populated groups \citep{2009ApJ...692..298W}, where mergers become important \citep{2014ApJ...782...53M}. In the following pages, we summarise the main points of the talk given in the IAU Symposium 321 in Toledo (Spain), in March 2016.
The talk and these proceedings were largely based on \citet{2015A&A...573A..78Q} and \citet{2015A&A...579L...2Q}.

\section{Bulge--disc coupling and pseudobulges in lenticular galaxies}

Analysing photometric decompositions of deep near-infrared observations, \citet{2010MNRAS.405.1089L} found strong scaling relations between the discs and bulges of lenticular galaxies. We have performed equivalent photometric decompositions on simulated major mergers, a subset 
of the GalMer database \citep{2010A&A...518A..61C}. 
We find that, under favourable orbital configurations, discs are first destroyed but then rebuilt short ($\sim$1\,Gyr) after the merging process is over. The possibility of disc survival in simulations of major mergers of spirals had already been implied by some authors \citep[e.g.~][]{2005ApJ...622L...9S, 2009ApJ...691.1168H,2016arXiv160203189A}, but we have taken one step forward and probed the potential merger origin of lenticular galaxies by comparing simulations with observations. For that, we created mock photometric images of the resulting relaxed galaxies using a variable, physically motivated mass-to-light ratio, and we performed photometric structural decompositions on the mock images of the merger remnants.
The conclusion is that the lenticular-looking galaxies emerging from those simulated major mergers obey the photometric scaling relations of real S0s. 

Figure\,\ref{fig:kinem} (left) illustrates the scaling relation between disc and bulge sizes in our simulation remnants compared to real lenticulars. For a given bulge radius, only some disc sizes are allowed by nature in lenticular galaxies, and the simulated major mergers can reproduce that bulge--disc coupling.
The simulation points cluster towards the top-right corner, 
but this is because the initial spiral models of GalMer are already very massive; starting from smaller progenitors would, presumably, reproduce the scaling relation over a larger range.
We have analysed eight additional photometric planes in \citet{2015A&A...573A..78Q}, including bulge-to-total ratios, magnitudes, and S\'ersic indices; we have also shown that these mergers can reproduce the pseudobulges detected in real lenticulars. In a companion study, we have also proved the existence of antitruncations in these merger remnants, which follow the same scaling relations as observations \citep{2014A&A...570A.103B}.

\section{Simultaneous change in concentration and angular momentum}
\label{lambda}

\begin{figure*}[t]
\begin{center}
\includegraphics[width=1.0\textwidth]{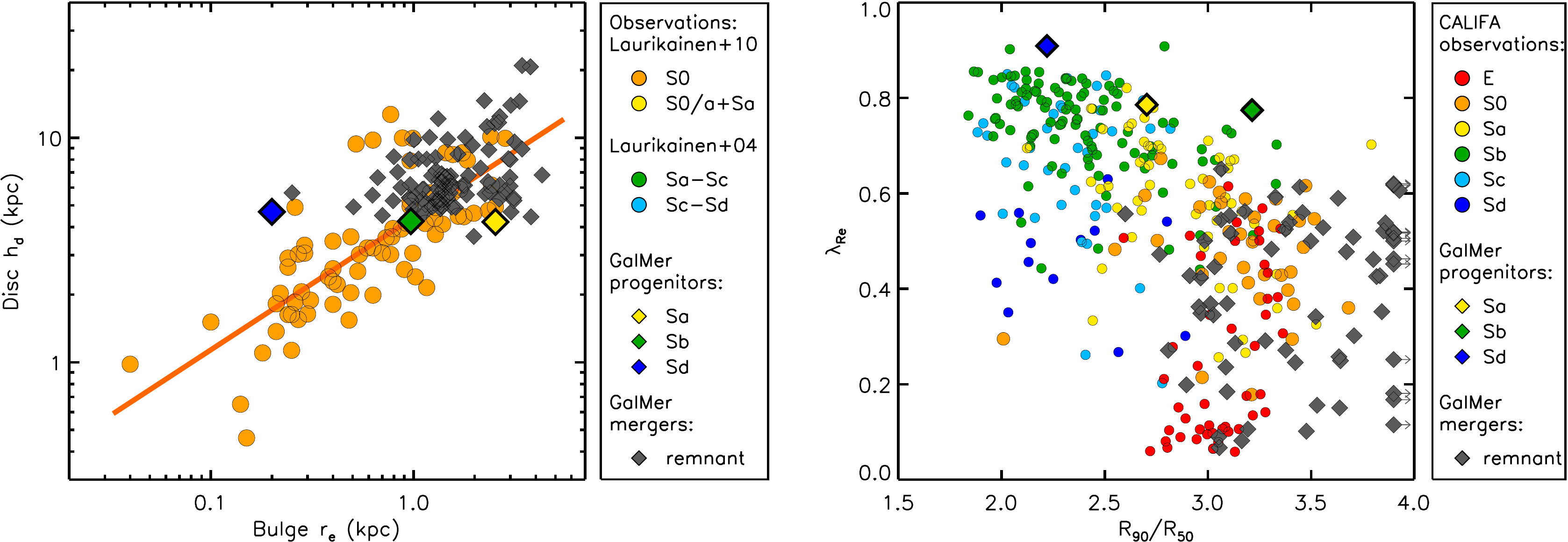} \caption{\textit{Left}: the major merger remnants (black diamonds) follow the same scaling relation as real lenticulars (orange circles) from \citet{2010MNRAS.405.1089L}. \textit{Right}: specific angular momentum ($\lambda_\mathrm{Re}$) versus concentration (\textit{R}$_{90}$/\textit{R}$_{50}$) for our major merger simulations (diamonds), and observational CALIFA datapoints. All parameters correspond to an edge-on view.}
\label{fig:kinem}
\end{center}
\end{figure*}

Late-type spirals seem to be incompatible with lenticular galaxies in a plane of angular momentum ($\lambda_\mathrm{Re}$) versus concentration (\textit{R}$_{90}$/\textit{R}$_{50}$), according to recent observational findings from the CALIFA team (\citealt{2015IAUS..311...78F}; van de Ven et al.~in preparation).  
Fig.\,\ref{fig:kinem} (right) reproduces this observational plane, and we overplot analogous measurements for the major merger progenitors and remnants that we introduced above. The simulated major mergers induce a simultaneous change in concentration and angular momentum, which can explain the stark offset between real spirals and lenticulars \citep{2015A&A...579L...2Q}; while early-type spirals (Sa) and lenticulars largely overlap, the change  between late-type spirals and S0s becomes more obvious. $\lambda_\mathrm{Re}$ is a proxy for the stellar angular momentum, calculated within the effective radius of the galaxy (\textit{R}$\leq$\textit{R}$_\mathrm{e}$). 
We estimate the light concentration through the Petrosian ratio \textit{R}$_{90}$/\textit{R}$_{50}$, measured on the 1D azimuthally averaged profile of our mock SDSS \textit{r} band images. The outliers in terms of concentration can be understood as the result of young stellar populations forming in the centre (the merger can funnel gas inwards), and because we are analysing those remnants short after coalescence ($\sim$1\,Gyr after the merger is completed), whereas most of the observed lenticulars have probably had much longer timescales to relax.
This analysis suggests that major mergers are able to transform spirals into realistic lenticular galaxies of lower $\lambda_\mathrm{Re}$ and higher concentration, in agreement with the observations from CALIFA. This observed evolution cannot be easily explained by  simple fading mechanisms such as ram pressure stripping, which are not expected to change the angular momentum of the initial galaxy significantly.

Summing up, we have shown that major mergers of spirals can sometimes rebuild discs, leading to lenticular galaxies in good agreement with photometric and kinematic observations.
We have also emphasised the importance of adopting a correct mass-to-light ratio when comparing such galaxy merger simulations with observations. Of course, our results do not necessarily imply that major mergers are the leading mechanisms in producing lenticular galaxies, and other processes could act in concert with mergers, or operate preferentially over certain environments; their relative relevance still remains to be quantified.


\bibliography{S0s.bib}{}
\bibliographystyle{aa}{}

\end{document}